\documentclass[a4paper,11pt]{article}
\usepackage{pos}

\title{WZ Sge-type dwarf novae: an extreme laboratory in cataclysmic variables}

\author*[a,b]{Yusuke Tampo}

\affiliation[a]{South African Astronomical Observatory\\ PO Box 9, Observatory, 7935, Cape Town, South Africa}

\affiliation[b]{Department of Astronomy, University of Cape Town\\ Private Bag X3, Rondebosch 7701, South Africa}

\emailAdd{yusuke@saao.ac.za}

\abstract{
WZ Sge-type dwarf novae form one of the most intriguing classes of compact accreting binaries. They are recognized as the most evolved population of hydrogen-rich cataclysmic variables. Yet they exhibit energetic disk-powered outbursts with an amplitude of 6–9 mag, duration of a month, and decade-long outburst cycles. Despite the dramatic increase in the number of WZ Sge stars over the last decades, there are many unresolved questions, both in terms of their binary evolution and outburst mechanism. In this proceeding paper, I review the recent studies on WZ Sge stars, focusing on their outbursts. 
Recent observations, both photometrically and spectroscopically, have found the absence of enhanced emission from the hotspot during the early outburst rise in WZ Sge stars, casting doubt on the occurrence of the enhanced mass transfer. Meanwhile, several WZ Sge stars are newly suggested to harbor a magnetic white dwarf, inferring an inner disk truncation in quiescence. Diversity among systems has been discovered: a WZ Sge star with superoutbursts both accompanying and lacking an early superhump phase, one possibly harboring an ONe and massive white dwarf, and one showing optical spectra strongly affected by disk winds. I introduce the connections of WZ Sge stars to period bouncers, helium-rich AM CVn stars, and low-mass X-ray binaries. Finally, prospects of WZ Sge stars in the upcoming time-domain surveys, such as the Rubin Observatory LSST, are presented.
}

\FullConference{87th Fujihara Seminar: The 50th Anniversary Workshop of the Disk Instability Model in Compact Binary Stars (DIM50TH2025)\\
22-26 September 2025\\
Tomakomai, Japan\\}



\begin{document}
\maketitle

\section{Introduction}

Among various subtypes of dwarf novae (DNe; see a review \cite{osa96review, kim20thesis}), WZ Sge-type DNe are characterized by their short orbital period ($P_\text{orb} \simeq$ 0.05--0.06 d), low mass-transfer rate ($\dot{M_\text{tr}} \sim 10^{15}$ g s$^{-1}$), and small mass ratio ($q\leq 0.1$), yet showing energetic accretion-powered outbursts with an outburst amplitude of 6--9 mag, outburst duration of 3--4 weeks, and outburst cycle of decades in optical \cite{kat15wzsge}. Indeed, the prototype star WZ Sge was initially classified as a recurrent nova due to its large amplitude and long outburst cycle. This view was altered around 1980, when various authors \cite{pat78wzsgeiauc3311, ort80wzsge, pat81wzsge} found that its photometric and spectroscopic properties in the 1978 outburst are consistent with DN outbursts rather than nova eruptions. This was exactly when the disk instability model (DIM) of DN outbursts had been established \cite{osa74DNmodel, mey81DNoutburst}. Soon after, various authors have pointed out that the simple DIM with a low mass-transfer rate cannot explain its outburst energetics (e.g., \cite{sma93wzsge}). 

\subsection{Optical light curves}

\begin{figure}[tbh]
\begin{center}
    \includegraphics[width=0.6\linewidth]{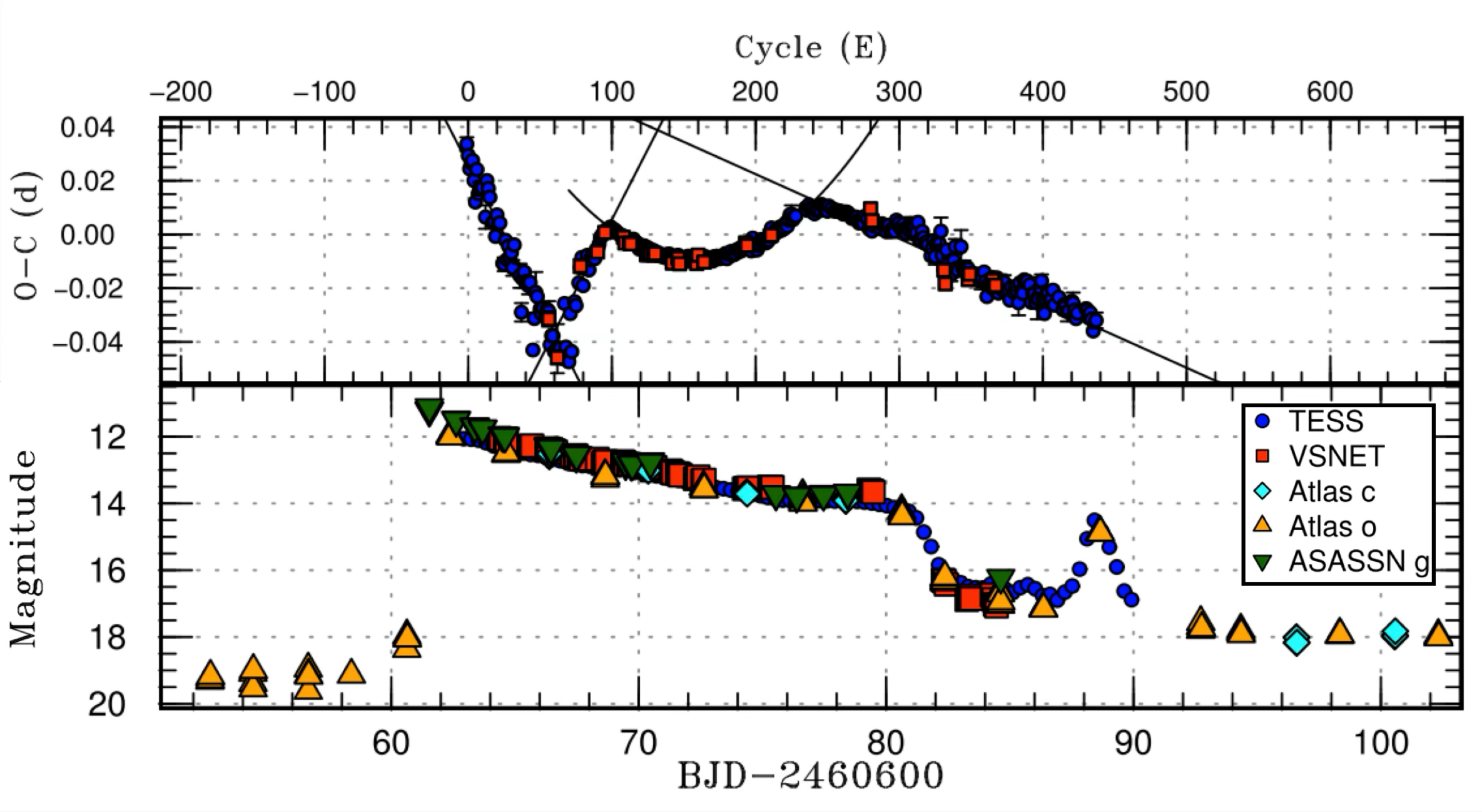}
    \includegraphics[width=0.35\linewidth]{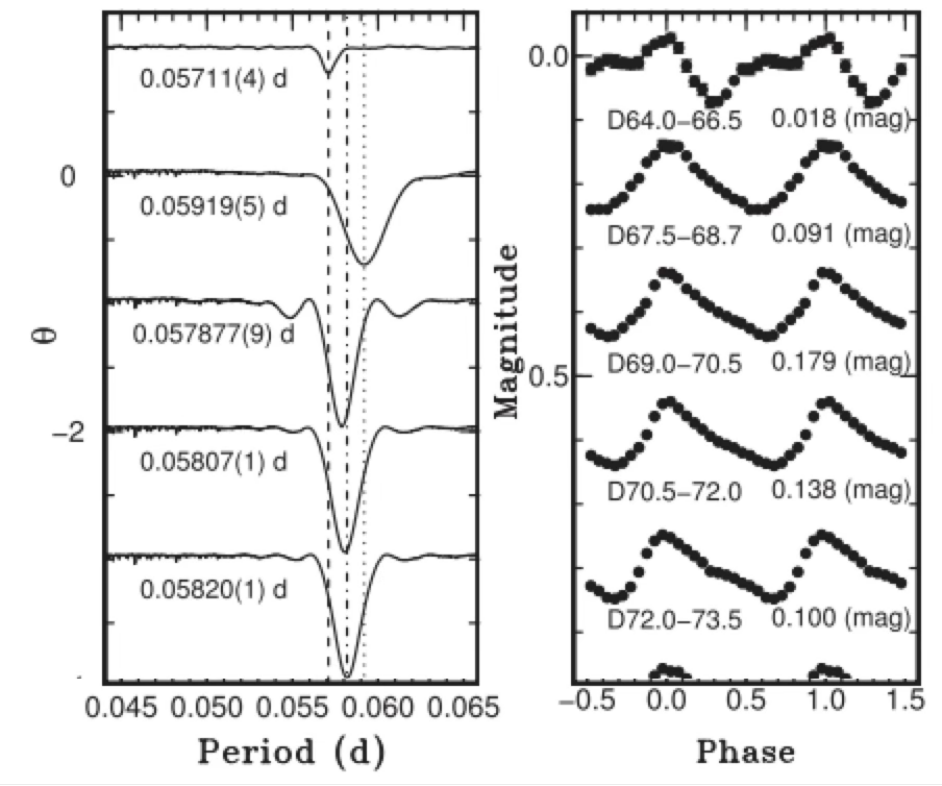}
    \caption{
            Left panels; the $O-C$ diagram of superhump maxima (top) and optical light curve (bottom) of a WZ Sge star ASASSN-24hd observed by TESS (blue circles) and through the VSNET collaboration (red squares; \cite{VSNET}). The solid lines in the $O-C$ diagram represent the periods of early, stage-A, stage-B, and stage-C superhumps. 
            Right panels: best periods in PDM analysis (left) and phase-averaged profiles of superhumps of different phases (right).
            Data is taken from \cite{tam25asassn24hd}.}
    \label{fig:OC-LC}
\end{center}
\end{figure}

As mentioned above, WZ Sge stars show the most energetic outbursts among DNe (Figure \ref{fig:OC-LC}).  All their outbursts are accompanied by small modulations with a period close to the orbital one, known as superhumps (right panels of Figure \ref{fig:OC-LC}). Based on the changes in superhump periods and profiles, a superoutburst in a WZ Sge star is divided into several phases.  Approximately in the first half of the superoutburst, double-peaked modulations with a period equal to the orbital one are observed, known as early superhumps. This is understood as a geometrical effect of the vertically extended double arm structure excited by the 2:1 tidal resonance, only in a system with a mass ratio lower than 0.08 \cite{lin79lowqdisk, osa02wzsgehump}. The detection of early superhumps is regarded as a criterion of a WZ Sge star in the current classification regime of DNe. They are followed by ordinary superhumps, with a single-peaked profile and a period a few percent longer than the orbital one, excited by the 3:1 tidal resonance \cite{whi88tidal}. These ordinary superhumps show essentially the same behavior as those in SU UMa-type DNe without the excitement of 2:1 resonance, explained in the thermal-tidal instability (TTI) model \cite{osa89suuma, Pdot}. 

\subsection{Discovery statistics}

\begin{figure}[tbh]
\begin{center}
    \includegraphics[width=0.35\linewidth]{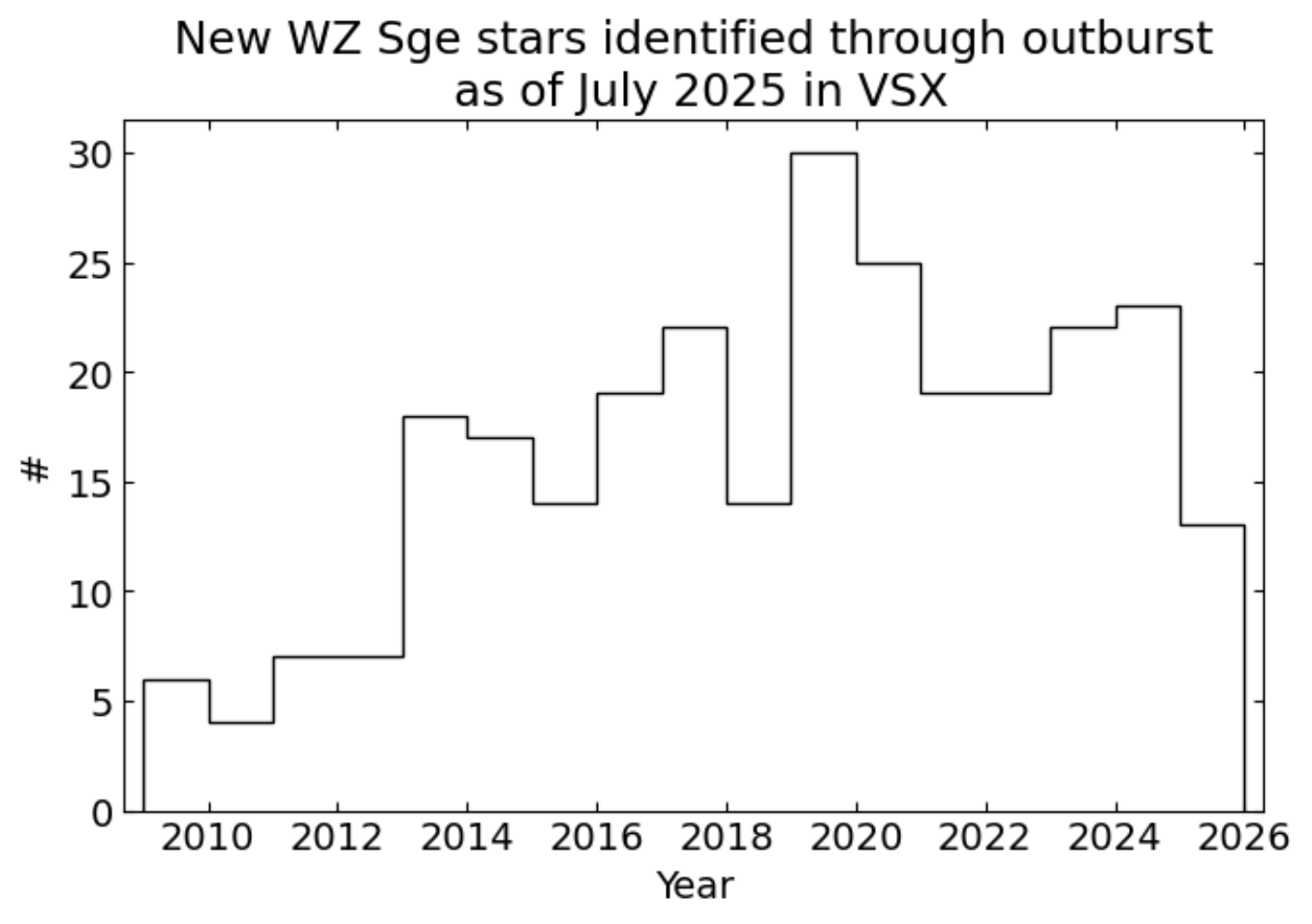}
    \includegraphics[width=0.63\linewidth]{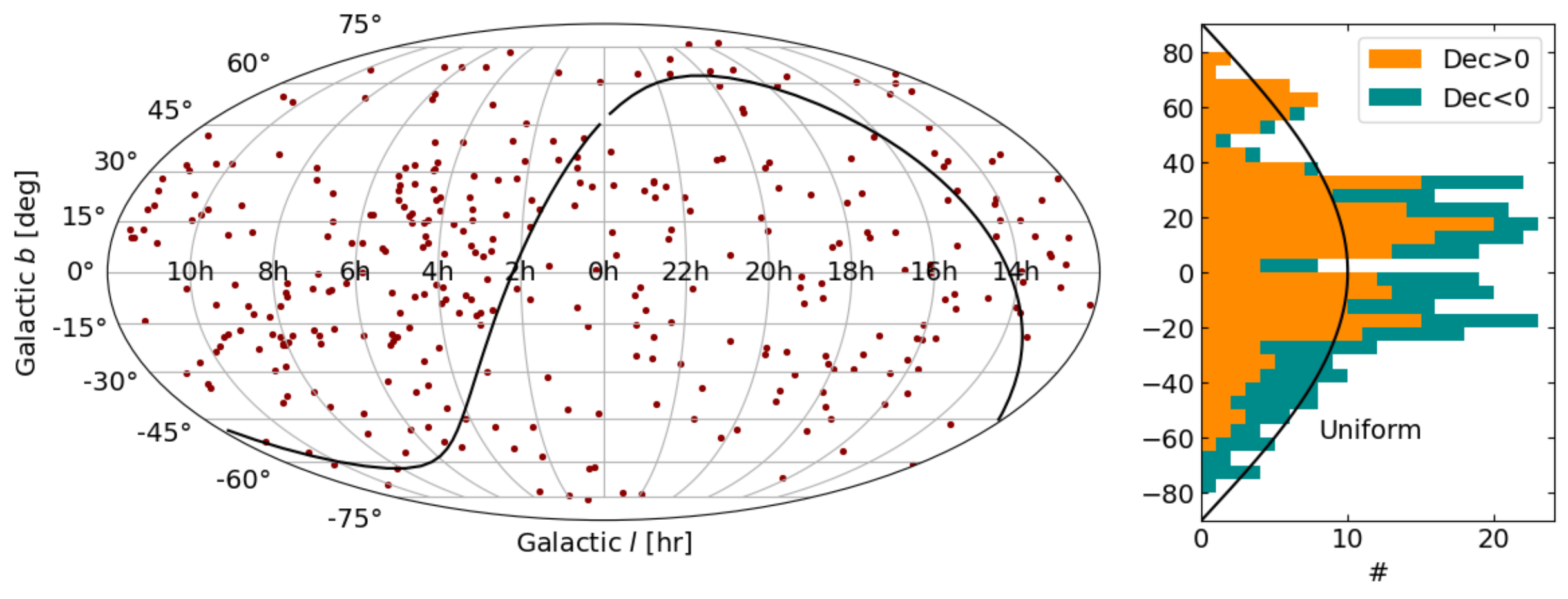}
    \caption{
            Left panel; the yearly number of new discoveries of WZ Sge stars through their outburst, according to VSX.
            Right panel; the sky distribution of WZ Sge stars in the Galactic coordinate. The black line on the right panel represents the uniform distribution on the sky sphere ($\propto \cos b $).}
    \label{fig:dist}
\end{center}
\end{figure}

As of July 2025, about 350 systems are registered as a WZ Sge star (or its candidate) in the AAVSO VSX \cite{VSX}. The left panel of Figure \ref{fig:dist} shows the annual number of new WZ Sge stars discovered through an outburst. This is limited by the performance of time-domain surveys and, more importantly, the limiting magnitude of small-aperture telescopes, which are used to determine the superhump properties. Hence, WZ Sge stars fainter than 16 mag at outburst maximum are largely missed. The right panels of Figure \ref{fig:dist} show the distribution of WZ Sge stars in the Galactic coordinate. This clearly indicates their Galactic disk origin.

\subsection{Binary evolution}

In addition to the importance of WZ Sge stars in the context of the DIM, they are a benchmark population among close binary systems, because they represent the most evolved state of low-mass binary systems. However, it has been pointed out that there is a large discrepancy between the volume density of CVs in models and observations \cite{pal20GaiaCVdensity}. The population models also predict that a large fraction (40--70\%) of Galactic CVs must evolve beyond the period minimum, which does not agree with the observations \cite{pal20GaiaCVdensity}. It is worth noting that since the period bounce happens around a mass ratio of 0.08 \cite{kni11CVdonor}, an outburst in a period bouncer system must be a WZ Sge-type DN outburst.
Superhump observations of WZ Sge stars are advantageous in studies of CV evolution because (1) early superhumps provide the orbital period, and (2) superhump excess over the orbital period is linked to the mass ratio \cite{pat05SH, kat13qfromstageA}. Moreover, these observations are done when a system is in outburst and bright, compared to the eclipse and radial velocity observations obtained in faint quiescence.

\section{Remaining questions on WZ Sge-type dwarf novae}

Besides the successful explanation of DN outbursts of SS Cyg stars and SU UMa stars with DIM, it has been found that the DIM simply lowering the mass-transfer rate may not reproduce the long faint quiescence and following energetic outbursts of WZ Sge stars. 
This is mainly because the total disk mass not triggering an outburst $M_{\rm disk, max}$ (Equation \ref{eq:1}; \cite{sma93wzsge}) and outburst recurrence time $t_{\rm rec}$ (Equation \ref{eq:2}; \cite{ham97wzsgemodel}) are described as a function of the disk viscosity in quiescence $\alpha_{\rm c}$ and the mass transfer rate $\dot{M}_{\rm tr}$.

\begin{equation}
    M_{\rm disk, max} = 5.6 \times 10^{21} {\alpha_{\rm c}}^{-0.79}~{\rm g} 
\label{eq:1}
\end{equation}

\begin{equation}
t_{\rm rec}  
\sim 4 \left( \frac{\alpha_{\rm c}}{10^{-2}} \right)^{-0.8} 
\left( 1 - \frac{\dot{M}_{\rm acc}}{\dot{M}_{\rm tr}} \right)^{-1}~{\rm yr} 
\label{eq:2}
\end{equation}

From these equations, it is clear that lowering just the mass-transfer rate (and accretion rate) results in outbursts with a longer outburst cycle but similar energetics to standard SU UMa stars, triggered at the inner disk due to a lower mass-transfer rate.
To solve this, two possibilities have been mainly proposed.
The first is the TTI model with very low disk viscosity in quiescence ($\alpha_{\rm c}\leq0.001$ compared to 0.01 in standard SU UMa stars) \cite{sma93wzsge, osa95wzsge}, as clear in the equations above. Another approach with the 'standard' quiescence viscosity ($\alpha_{\rm c} \simeq 0.01$) is the truncation of the inner disk by either the magnetosphere of the white dwarf (WD) or the evaporation effect in quiescence \cite{las95wzsge, ham97wzsgemodel}, which prevents the matter from inflowing into the inner disk and stabilizes the disk. These authors also claim that, in this case, an outburst in WZ Sge stars must be triggered and maintained by the enhanced mass transfer from the secondary star, because such a truncated disk only contains disk mass an order of magnitude less than that required to trigger an outburst. Some works employ a hybrid approach with a low quiescence viscosity, inner disk truncation, and constant mass transfer rate \cite{mey98wzsge, mat07wzsgepropeller}. Here in this section, I review some recent works that provide insight into understanding their outburst scenarios and diversity.

\subsection{What enables a long outburst cycle?}

\begin{figure}[tbh]
\begin{center}
    \includegraphics[width=0.6\linewidth]{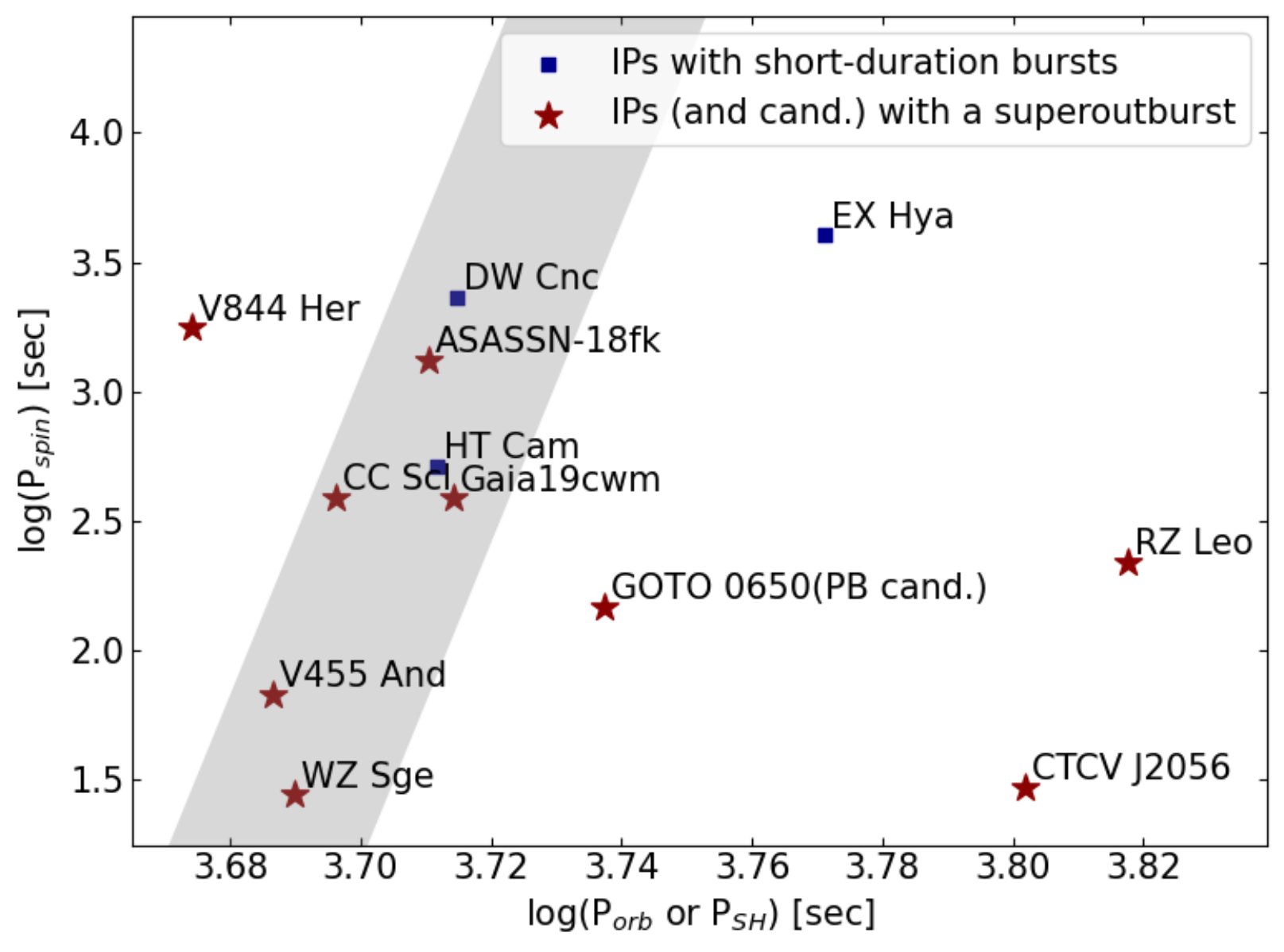}
    \caption{
            The orbital periods vs the spin periods of short-$P_\text{orb}$ CVs with burst-like natures \cite{wou12ccscl, szk17ccsclrzleo, pav19a19fk, lop20j2056, gre23v844her, kol25gaia19cwm, cas25goto0650}. The stars represent a system with a superoutburst (i.e., triggered by thermal instability), while squares represent one with short-lived bursts which are not likely triggered by disk instability. PB stands for a period bouncer candidate. The gray area roughly represents the proposed evolutionary path on this diagram \cite{pav19a19fk}.}
    \label{fig:IPs}
\end{center}
\end{figure}

As mentioned above, a large magnetosphere of the magnetized WD is one of the proposed mechanisms of the inner disk truncation. 
Although the intermediate-polar (IP) nature of WZ Sge itself was initially proposed in the 1980s \cite{pat80wzsge}, it remains unclear if this is a universal characteristic of the subtype. Thanks to the development of observation facilities, the number of short-$P_\text{orb}$ CVs with the detection of the (possible) WD spin period has increased dramatically over the last decade \cite{wou12ccscl, szk17ccsclrzleo, pav19a19fk, lop20j2056, gre23v844her, kol25gaia19cwm, cas25goto0650}.
Figure \ref{fig:IPs} shows the orbital and spin periods of these samples. Although once an empirical evolutionary path of the orbital and spin periods has been suggested (gray area in Figure \ref{fig:IPs}; \citep{pav19a19fk}), it is clear that they widely spread around on this diagram, suggesting a diverse nature of their inner disk truncation, rather than uniform. In addition, QPOs observed in some WZ Sge stars in quiescence are also attributed to the magnetic nature of the WD \cite{ver24awdqpo}. It must be noted that, however, this is not a comprehensive view of the magnetic WDs in short-$P_\text{orb}$ CVs at all, as the detection and confirmation of a spin period require high-quality data for a long enough baseline.

In the framework of DIM, an accretion disk must accumulate its mass over the years in quiescence. A possible evidence of this has been found in V3101 Cyg, one of the closest WZ Sge stars, which showed a gradual brightening ($\simeq - 0.1$ mag year$^{-1}$) for five years before an outburst \cite{tam20j2104}. Since the typical quiescence brightness of WZ Sge stars is 20 mag or fainter, future time-domain surveys such as LSST will be capable of testing whether this is a universal trend in WZ Sge stars.

\subsection{How does a system undergo an outburst?}

One of the natural expectations from the increase in the mass transfer rate is the enhanced emission from the hot spot. Thanks to the continuous coverage of TESS, the early rise of the outburst of V748 Hya was well traced, reporting the non-detection of the hot spot at the level of $\leq 1 \times 10^{31}$ erg s$^{-1}$ \cite{tam25wzsgetess}. This limits the corresponding mass-transfer rate to be below $1 \times 10^{16}$ g s$^{-1}$. This value is well smaller than the expected mass-transfer rate $\simeq1 \times 10^{18} (10^{19})$ g s$^{-1}$ to transfer $10^{24}$ g over 10 (1) days, giving a strong constraint on the occurrence of the mass transfer burst at the beginning of an outburst. Another constraint is provided by the coincident time-resolved spectroscopy on the rise of the 2007 superoutburst of V455 And \cite{tov22v455andspec}, which does not report any evidence of the enhanced hotspot in their Doppler maps during the outburst rise.

The well-observed Kepler and TESS light curves show a double-powerlaw rise in the flux scale \cite{rid19j165350, tam25wzsgetess}, broken when the system is just $\simeq1\%$ brightness of the outburst maximum. Although its physical origin is yet unclear, the simulation work also shows a break at a similar flux level and timing, which is attributed to the start of the heating wave propagating across the entire disk \cite{jor24ttisimulation}.

\subsection{What explains their diverse outbursts?}

The modifications of either a lower quiescence viscosity or an inner disk truncation to the original DIM were initially proposed to explain the outbursts of WZ Sge itself. However, with hundreds of WZ Sge stars known today, a comprehensive view of their diversity is a key to deepening our understanding of the DIM.

\subsubsection{Borderline between SU UMa stars and WZ Sge stars}

In the classification scheme of CVs, WZ Sge stars are categorized as a special subtype of SU UMa stars, characterized by the lowest mass ratios. However, it remains unclear if there is a clear borderline between standard SU UMa stars and WZ Sge stars. For example, several WZ Sge stars with a mass ratio presumably larger than 0.1 have been reported \cite{wak17asassn16eg}.
Moreover, various authors \cite{kim16alcom, tam20j2104, tam23v627peg, tam25wzsgetess} have found some WZ Sge stars that exhibit superoutbursts not only accompanied by the early superhump phase but also lacking one. The latter superoutbursts are characterized by smaller outburst amplitude and shorter outburst duration, observed within a few years after the main superoutburst with early superhumps. This fact proves that, although the mass ratio determines whether a system can excite the 2:1 resonance, the disk mass at the onset of an outburst is a more significant factor in determining whether the disk reaches the 2:1 resonance radius or not in a respective outburst.

\subsubsection{MASTER OT J030227.28$+$191754.5}

\begin{figure}[tbh]
\begin{center}
    \includegraphics[width=0.50\linewidth]{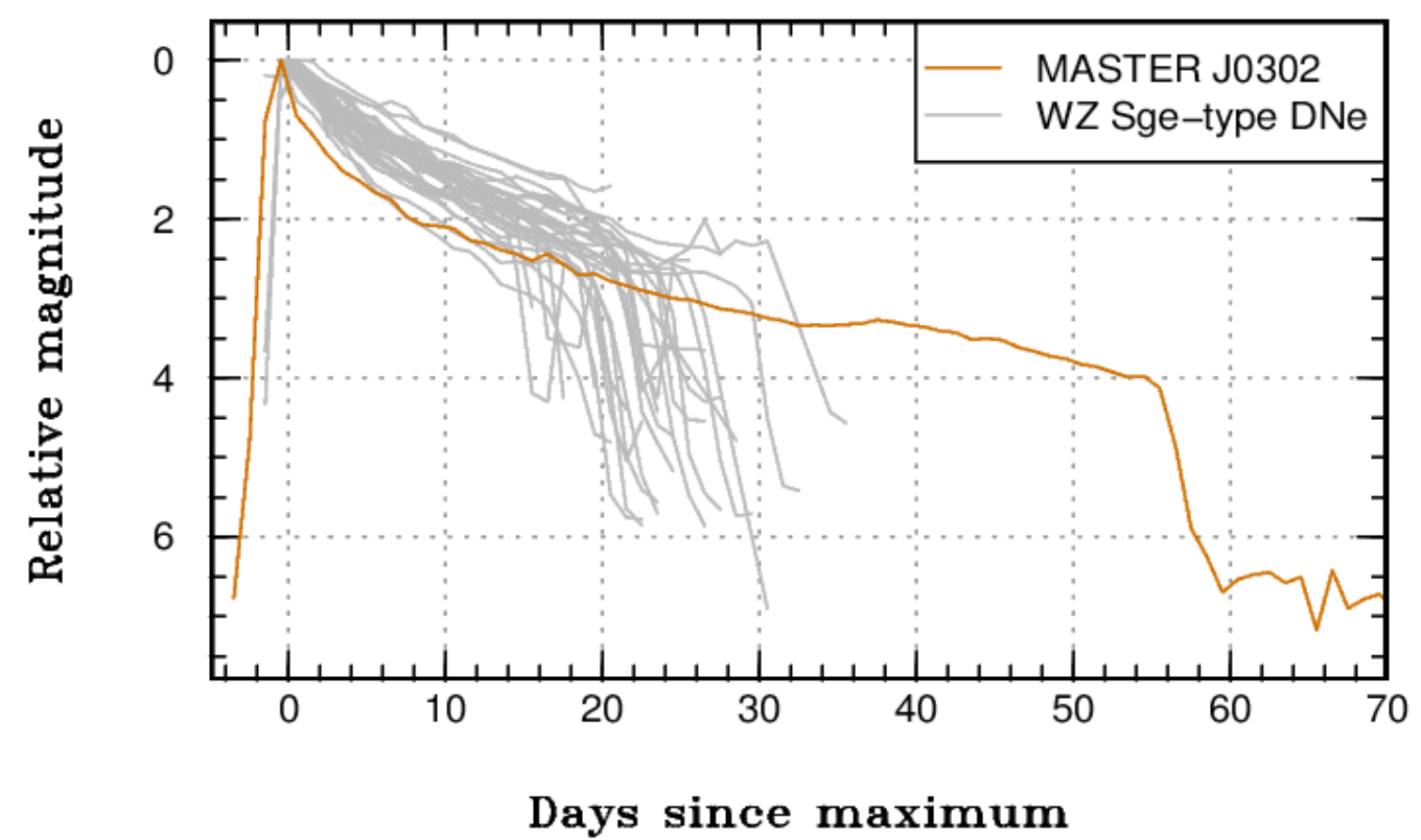}
    \includegraphics[width=0.47\linewidth]{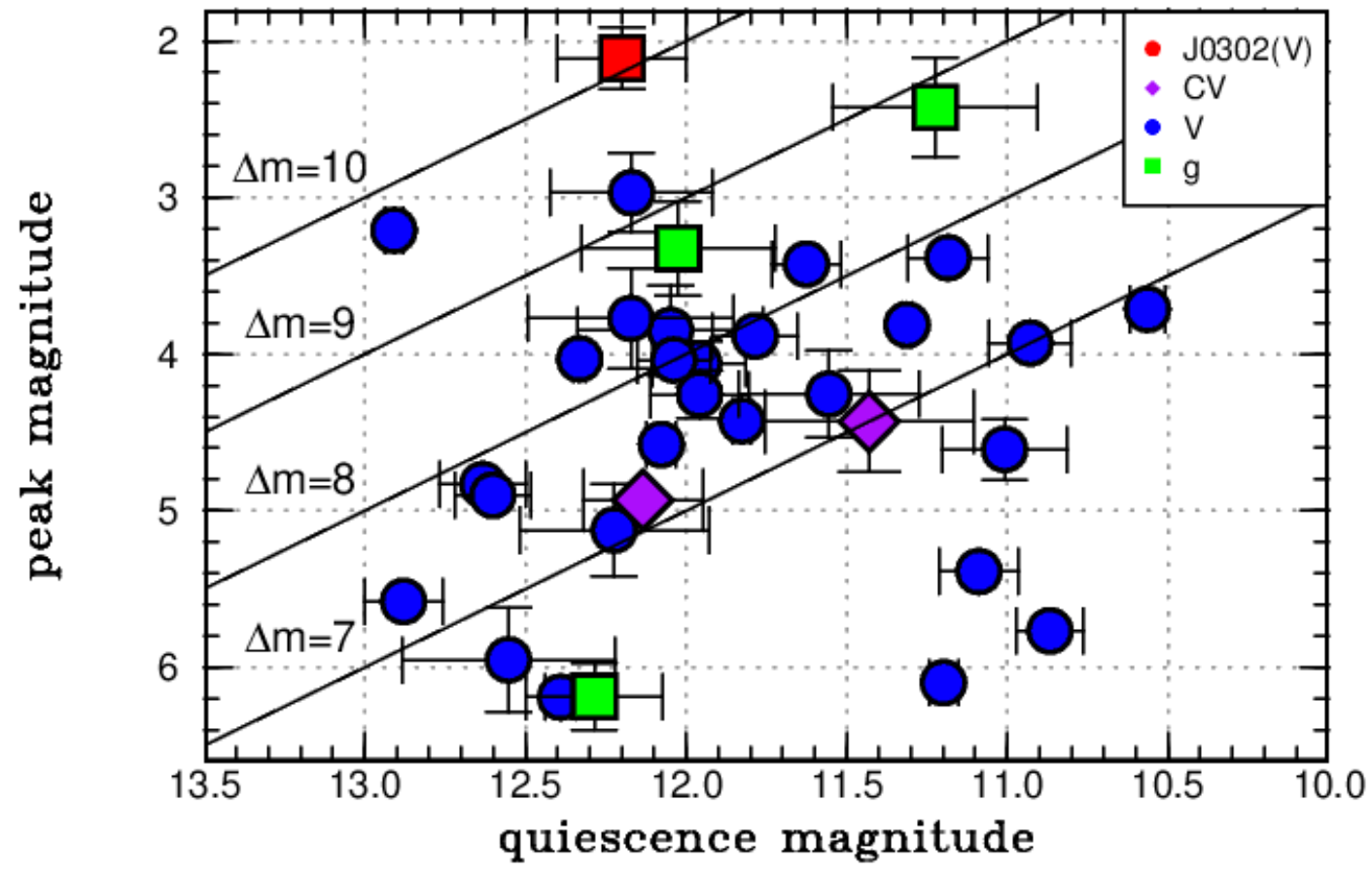}
    \caption{
            Left panel; the optical light curve comparison of J0302 (orange) and other WZ Sge stars (gray), normalized at the outburst maxima. 
            Right panel; the absolute magnitudes in quiescence and at the outburst maximum of WZ Sge stars. The solid lines represent the outburst amplitudes of 7--10 mag.
            }
    \label{fig:j0302}
\end{center}
\end{figure}

Among the known WZ Sge stars, MASTER OT J030227.28$+$191754.5 (hereafter J0302) showed a very distinctive light curve of the outburst \cite{kim23j0302, tam24j0302}, with 10.2-mag amplitude and 60-d duration. Figure \ref{fig:j0302} compares it to other WZ Sge stars in its optical light curve (left) and absolute magnitudes (right), making it the most energetic DN outburst ever. However, its orbital period and mass ratio are well within the range of WZ Sge stars, and its superhump properties agree with the TTI model. 
Moreover, its X-ray spectrum at the outburst maximum exhibits a blackbody component with a luminosity of $\simeq 10^{34}$ erg s$^{-1}$ and a temperature of $\simeq$30 eV. By interpreting that this blackbody comes from the optically-thick boundary layer around the WD, its WD mass is estimated as 1.15--1.34 $M_\odot$ \cite{kim23j0302}. Hence, J0302 is the first CV system with a massive ONe WD around the period minimum. Moreover, in the most massive case, the total mass in the binary exceeds the Chandrasekhar mass, proposing a new evolutionary path to the accretion-induced collapse of an ONe WD. It was discussed that, however, this massive WD itself may not fully explain the observed energetics of the outburst, but even lower quiescence viscosity than other WZ Sge stars may be required in the context of the TTI model \cite{tam24j0302}.

\subsubsection{Disk wind laboratory}

\begin{figure}[tbh]
\begin{center}
    \includegraphics[width=0.58\linewidth]{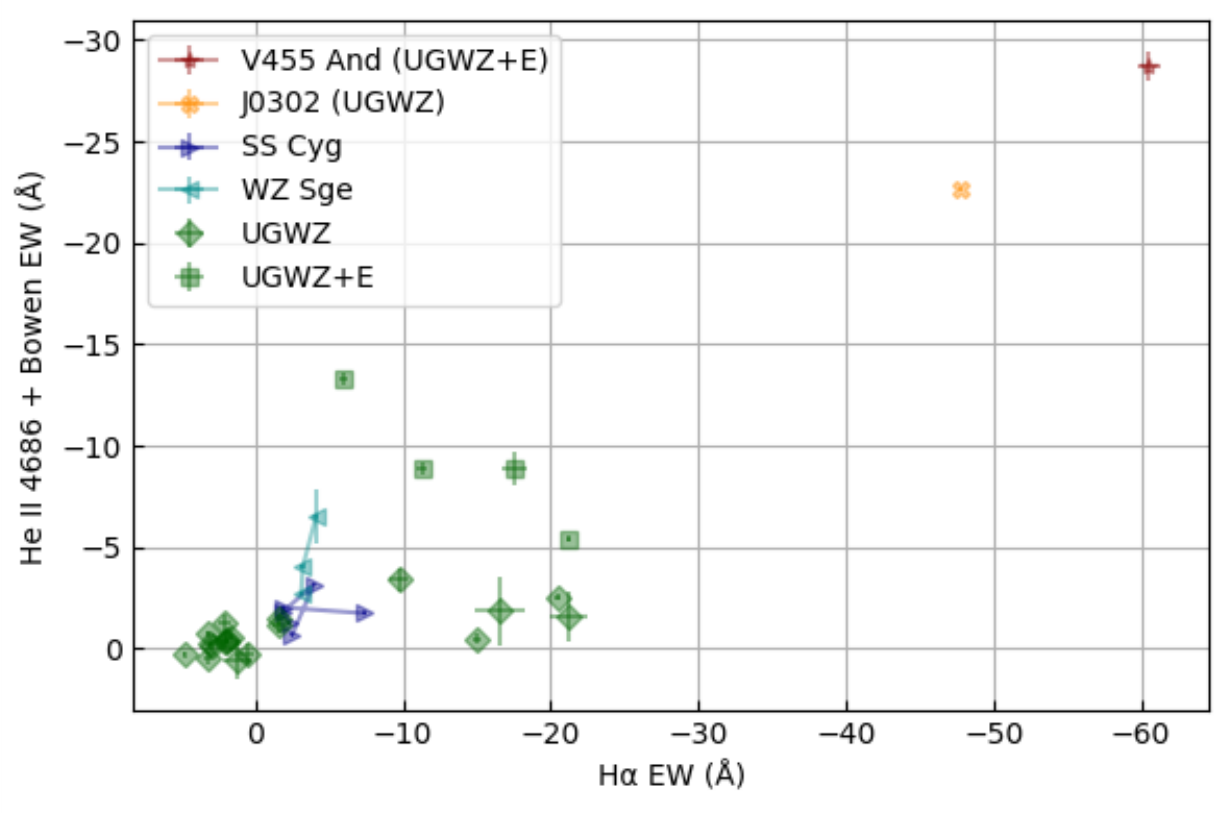}
    \includegraphics[width=0.38\linewidth]{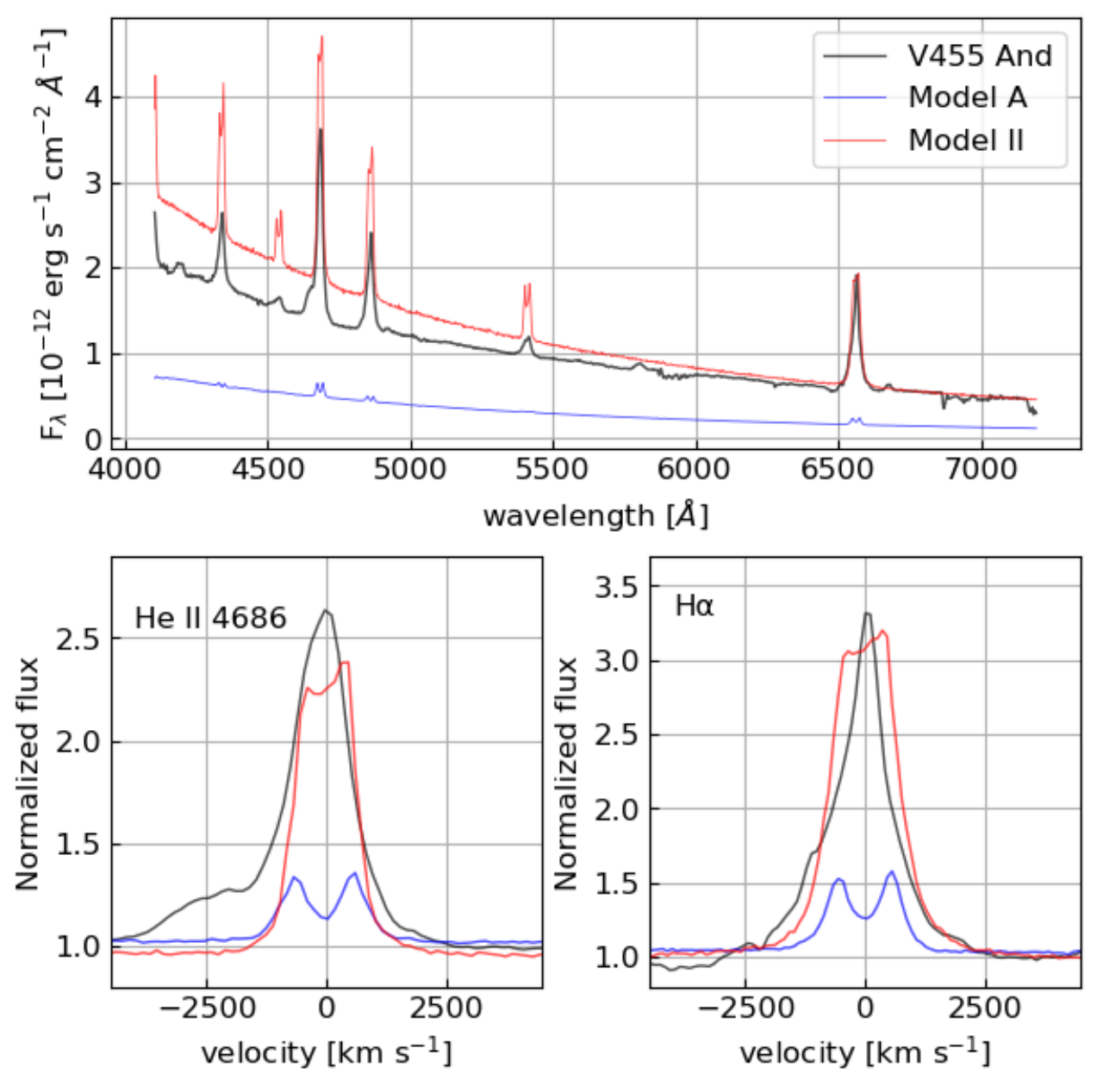}
    \caption{
            Left panel; the EWs of H$\alpha$ and He~\textsc{ii} 4686 around the outburst maximum of WZ Sge stars (references in \cite{tam21seimeiCVspec}, \cite{tam22v455andspec, tam24j0302} and ATel reports with the observations from the Seimei telescope).
            Right panel; the observed spectrum of V455 And at its outburst maximum (black; \cite{tam22v455andspec}) and simulated spectra with disk wind models (blue and red; \cite{tam24dnwind}).
            }
    \label{fig:wind}
\end{center}
\end{figure}

Outflows in compact binaries, such as disk winds and jets, may also regulate disk instability and binary evolution \cite{can88angularmomentumloss, tet18winddim}. Our understanding of disk winds in disk-dominated CVs has been established mainly by UV spectra, which contain P-Cygni features in some resonance lines \cite{lon02python}. On the other hand, the disk atmosphere is attributed to the main source of optical emission lines in the majority of WZ Sge stars in outburst \cite{tam21seimeiCVspec}. The left panel of Figure \ref{fig:wind} represents the equivalent widths (EWs) of H$\alpha$ and He~\textsc{II} 4686 of WZ Sge stars around the outburst maximum, mostly observed with the Seimei telescope in Okayama, Japan \cite{kur20seimei}. The systems with H$\alpha$ in absorption (positive EW) and no He~\textsc{II} can be explained by a face-on optically-thick disk. Some systems with moderate emission of He~\textsc{II} are in fact an eclipsing system, and show a double-peaked profile of emission lines. Thus, emission lines likely originate from the rotating disk atmosphere. There are two systems with distinctively strong emission lines in both Balmer and He~\textsc{II} (V455 And and J0302) \cite{tam22v455andspec, tam24j0302}. Since their line profiles are dominated by the component narrower than the expectation from a Keplerian disk (right panels of Figure \ref{fig:wind}), their origin seems to differ from other systems. 

A simulation effort has been performed to reproduce these peculiar spectral features (colored lines in the right panels of Figure \ref{fig:wind}; \cite{tam24dnwind}). These authors find that the disk wind model (Model A in the figure), which has been developed to explain the UV spectral features \cite{lon02python, mat25sirocco}, can reasonably reproduce the observed spectrum of V455 And, once a disk accretion rate an order of magnitude higher than previous studies is applied (Model II in the figure). Such a high accretion rate of order of $\dot{M_\text{acc}} \simeq 10^{-7} M_\odot$ yr$^{-1}$ is implied from their bright optical continuum. Thus, high accretion rates are presumably the main cause of the significant appearance of disk wind signatures in these intrinsically bright systems.

\subsection{Application to other accreting binaries}

In addition to the intriguing nature of WZ Sge stars themselves, physics studied in WZ Sge stars has been applied to other compact binary systems. One such example is identifying a period bouncer candidate based on the outburst light curve, which is essentially a WZ Sge star with the lowest mass ratios. As many authors have pointed out (e.g. \cite{pat05SH, Pdot}), superoutburst and superhump properties are strongly linked to the mass ratio of a system. For example, the dip in the middle of the superoutburst plateau (type-E rebrightening episode), as well as negative $P_\text{dot}$ in stage-B ordinary superhumps and slow decline timescale in the ordinary superhump phase, are regarded as the smoking-gun features of a superoutburst in a period bouncer system \cite{kat13j1222, kim18asassn16dt16hg}.
The mass ratio becomes lower than 0.1 for helium-rich AM CVn stars with an orbital period longer than $\simeq$30 min \cite{gre25amcvns}, which enables the excitation of the 2:1 tidal resonance and WZ Sge-like behavior. Indeed, early superhumps and superhump properties, the same as hydrogen-rich WZ Sge stars, have been observed in some AM CVn systems \cite{iso19nsv1440, riv22a21au, tar25j0722}. 
Low-mass X-ray binaries (LMXBs) are not an exception. Both in WZ Sge stars and the LMXB MAXI J1820$+$70, the change of X-ray properties at the transition of the superhump stages has been observed \cite{neu18j1222gwlib, nii21a18ey}. These authors interpret that this change in X-ray reflects the geometrical change of the inner disk by the propagation of eccentricity.

\section{Bright future of WZ Sge-type dwarf novae}

\subsection{New catalogs of WZ Sge star candidates}

It has been almost half a century since WZ Sge was confirmed as a DN system in its 1978 superoutburst, and even longer since its identification as an accreting WD binary (e.g., \cite{krz64wzsge}). However, the vast majority of new WZ Sge stars are still discovered through their superoutbursts, and little is known about their quiescence and dormant systems. This has been changed thanks to Gaia and its parallax measurements. Combining with eROSITA and SDSS \cite{mun24erositacv, ini25sdsscvs}, these authors compiled the list of new accreting WDs, including some candidates around or even below the period minimum. Long-term observations by time-domain surveys are expected to detect their infrequent outbursts, and indeed some are observed. Their superhumps in outburst offer a great opportunity to validate their binary parameters, such as orbital period and mass ratio.

\subsection{WZ Sge stars in the upcoming instruments; LSST and LISA}

The upcoming facilities and instruments are also expected to discover many new CVs and to improve our understanding of WZ Sge stars. The Rubin Observatory LSST is about to start its operation at the end of 2025. The mock observation simulations predict the detection of more than one thousand outbursts from WZ Sge stars every year down to 24 mag by LSST \cite{buc25cvinlsst}. These authors also claim that LSST will provide a unique test case for the DIM and CV evolution models in globular clusters and Magellanic clouds, which have different star formation histories and environments compared to the solar neighborhood. Along with AM CVn stars, some WZ Sge stars are considered to be detected by the future space-based interferometer for gravitational waves like LISA \cite{sca23cvsinlisa, pog24cvgw}.

\subsection{Brightest WZ Sge stars}

Lastly, the brightest systems should not be forgotten. The outburst cycle of the prototype WZ Sge varies between 23 to 33 years. Because it has already been 24 years since its last outburst in 2001, it may undergo a new outburst anytime now. GW Lib is expected to outburst in the 2030s based on its previous two outbursts in 1983 and 2007. It has also been more than 10 years since the last outburst of the bright systems V455 And in 2007 and BW Scl in 2011. Outbursts in these systems will provide the best observation opportunity to test various remaining questions above with cutting-edge multi-wavelength facilities. I finally stress that classical time-resolved observations of superhumps from both professional and amateur astronomers are also strongly encouraged.

\section*{Acknowledgment}

I thank the organizers of the 87th Fujihara Seminar "The 50th Anniversary Workshop of the Disk Instability Model in Compact Binary Stars" for the invitation and the Fujihara Foundation for the travel support.  I am grateful to all my collaborators, especially D. Nogami, T. Kato, N. Kojiguchi, K. Isogai, M. Kimura, C. Knigge, and D.A.H. Buckley. 
I acknowledge all observers and observatories who have shared their valuable data with the Variable Star Network (VSNET; \cite{VSNET}), the Variable Star Observers League in Japan (VSOLJ\footnote{\url{http://vsolj.cetus-net.org/index.html}}), the American Association of Variable Star Observers (AAVSO\footnote{\url{https://www.aavso.org/}}), and other public database of variable stars.


\bibliographystyle{JHEP}
\bibliography{cvs} 

\newcommand{\noop}[1]{}

\providecommand{\href}[2]{#2}\begingroup\raggedright\begin{thebibliography}{10}

\bibitem{osa96review}
Y.~{Osaki}, \emph{{Dwarf-Nova Outbursts}}, \href{https://doi.org/10.1086/133689}{\emph{\pasp} {\bfseries 108} (1996) 39}.

\bibitem{kim20thesis}
M.~{Kimura}, \emph{{Observational and Theoretical Studies on Dwarf-nova Outbursts}}, Springer Theses, Springer Singapore, 1~ed. (2020), \href{https://doi.org/10.1007/978-981-15-8912-6}{10.1007/978-981-15-8912-6}.

\bibitem{kat15wzsge}
T.~{Kato}, \emph{{WZ Sge-type dwarf novae}}, \href{https://doi.org/10.1093/pasj/psv077}{\emph{\pasj} {\bfseries 67} (2015) 108} [\href{https://arxiv.org/abs/1507.07659}{{\ttfamily 1507.07659}}].

\bibitem{pat78wzsgeiauc3311}
J.~{Patterson}, J.T.~{McGraw}, R.E.~{Nather} and R.~{Stover}, \emph{{WZ Sagittae}}, {\emph{\iaucirc} {\bfseries 3311} (1978) 1}.

\bibitem{ort80wzsge}
S.~{Ortolani}, P.~{Rafanelli}, L.~{Rosino} and A.~{Vittone}, \emph{{The recent outburst of the dwarf nova WZ Sge.}}, {\emph{\aap} {\bfseries 87} (1980) 31}.

\bibitem{pat81wzsge}
J.~{Patterson}, J.T.~{McGraw}, L.~{Coleman} and J.L.~{Africano}, \emph{{A photometric study of the dwarf nova WZ SGE in outburst.}}, \href{https://doi.org/10.1086/159236}{\emph{\apj} {\bfseries 248} (1981) 1067}.

\bibitem{osa74DNmodel}
Y.~{Osaki}, \emph{{An Accretion Model for the Outbursts of U Geminorum Stars}}, \href{https://doi.org/10.1093/pasj/26.4.429}{\emph{\pasj} {\bfseries 26} (1974) 429}.

\bibitem{mey81DNoutburst}
F.~{Meyer} and E.~{Meyer-Hofmeister}, \emph{{On the elusive cause of cataclysmic variable outbursts.}}, {\emph{\aap} {\bfseries 104} (1981) L10}.

\bibitem{sma93wzsge}
J.~{Smak}, \emph{{WZ SGE as a Dwarf Nova}}, {\emph{\actaa} {\bfseries 43} (1993) 101}.

\bibitem{lin79lowqdisk}
D.N.C.~{Lin} and J.~{Papaloizou}, \emph{{Tidal torques on accretion discs in binary systems with extreme mass ratios.}}, \href{https://doi.org/10.1093/mnras/186.4.799}{\emph{\mnras} {\bfseries 186} (1979) 799}.

\bibitem{osa02wzsgehump}
Y.~{Osaki} and F.~{Meyer}, \emph{{Early humps in WZ Sge stars}}, \href{https://doi.org/10.1051/0004-6361:20011744}{\emph{\aap} {\bfseries 383} (2002) 574} [\href{https://arxiv.org/abs/astro-ph/0112309}{{\ttfamily astro-ph/0112309}}].

\bibitem{whi88tidal}
R.~{Whitehurst}, \emph{{Numerical simulations of accretion discs - I. Superhumps : a tidal phenomenon of accretion discs.}}, \href{https://doi.org/10.1093/mnras/232.1.35}{\emph{\mnras} {\bfseries 232} (1988) 35}.

\bibitem{osa89suuma}
Y.~{Osaki}, \emph{{A model for the superoutburst phenomenon of SU Ursae MAjoris stars.}}, {\emph{\pasj} {\bfseries 41} (1989) 1005}.

\bibitem{Pdot}
T.~{Kato}, A.~{Imada}, M.~{Uemura}, D.~{Nogami}, H.~{Maehara}, R.~{Ishioka} et~al., \emph{{Survey of Period Variations of Superhumps in SU UMa-Type Dwarf Novae}}, \href{https://doi.org/10.1093/pasj/61.sp2.S395}{\emph{\pasj} {\bfseries 61} (2009) S395} [\href{https://arxiv.org/abs/0905.1757}{{\ttfamily 0905.1757}}].

\bibitem{VSNET}
T.~{Kato}, M.~{Uemura}, R.~{Ishioka}, D.~{Nogami}, C.~{Kunjaya}, H.~{Baba} et~al., \emph{{Variable Star Network: World Center for Transient Object Astronomy and Variable Stars}}, \href{https://doi.org/10.1093/pasj/56.sp1.S1}{\emph{\pasj} {\bfseries 56} (2004) S1} [\href{https://arxiv.org/abs/astro-ph/0310209}{{\ttfamily astro-ph/0310209}}].

\bibitem{tam25asassn24hd}
Y.~{Tampo}, N.~{Kojiguchi}, T.~{Kato}, M.~{Kimura}, D.A.H.~{Buckley}, B.~{Monard} et~al., \emph{{ASASSN-24hd: A dwarf nova bridging WZ Sge-type and SU UMa-type superoutbursts}}, \href{https://doi.org/10.1093/pasj/psaf051}{\emph{\pasj} {\bfseries 77} (2025) 823} [\href{https://arxiv.org/abs/2504.20783}{{\ttfamily 2504.20783}}].

\bibitem{VSX}
C.L.~{Watson}, A.A.~{Henden} and A.~{Price}, \emph{{The International Variable Star Index (VSX)}}, {\emph{Society for Astronomical Sciences Annual Symposium} {\bfseries 25} (2006) 47}.

\bibitem{pal20GaiaCVdensity}
A.F.~{Pala}, B.T.~{G{\"a}nsicke}, E.~{Breedt}, C.~{Knigge}, J.J.~{Hermes}, N.P.~{Gentile Fusillo} et~al., \emph{{A Volume-limited Sample of Cataclysmic Variables from Gaia DR2: Space Density and Population Properties}}, \href{https://doi.org/10.1093/mnras/staa764}{\emph{\mnras} {\bfseries 494} (2020) 3799} [\href{https://arxiv.org/abs/1907.13152}{{\ttfamily 1907.13152}}].

\bibitem{kni11CVdonor}
C.~{Knigge}, I.~{Baraffe} and J.~{Patterson}, \emph{{The Evolution of Cataclysmic Variables as Revealed by Their Donor Stars}}, \href{https://doi.org/10.1088/0067-0049/194/2/28}{\emph{\apjs} {\bfseries 194} (2011) 28} [\href{https://arxiv.org/abs/1102.2440}{{\ttfamily 1102.2440}}].

\bibitem{pat05SH}
J.~{Patterson}, J.~{Kemp}, D.A.~{Harvey}, R.E.~{Fried}, R.~{Rea}, B.~{Monard} et~al., \emph{{Superhumps in Cataclysmic Binaries. XXV. q$_{crit}$, $\epsilon$(q), and Mass-Radius}}, \href{https://doi.org/10.1086/447771}{\emph{\pasp} {\bfseries 117} (2005) 1204} [\href{https://arxiv.org/abs/astro-ph/0507371}{{\ttfamily astro-ph/0507371}}].

\bibitem{kat13qfromstageA}
T.~{Kato} and Y.~{Osaki}, \emph{{New Method of Estimating Binary's Mass Ratios by Using Superhumps}}, \href{https://doi.org/10.1093/pasj/65.6.115}{\emph{\pasj} {\bfseries 65} (2013) 115} [\href{https://arxiv.org/abs/1307.5588}{{\ttfamily 1307.5588}}].

\bibitem{ham97wzsgemodel}
J.-M.~{Hameury}, J.-P.~{Lasota} and J.-M.~{Hure}, \emph{{A model for WZ SGE with `standard' values of alpha}}, \href{https://doi.org/10.1093/mnras/287.4.937}{\emph{\mnras} {\bfseries 287} (1997) 937} [\href{https://arxiv.org/abs/astro-ph/9701164}{{\ttfamily astro-ph/9701164}}].

\bibitem{osa95wzsge}
Y.~{Osaki}, \emph{{A Model for WZ Sagittae-Type Dwarf Novae: SU UMa/WZ Sge Connection}}, \href{https://doi.org/10.1093/pasj/47.1.47}{\emph{\pasj} {\bfseries 47} (1995) 47}.

\bibitem{las95wzsge}
J.P.~{Lasota}, J.M.~{Hameury} and J.M.~{Hure}, \emph{{Dwarf novae at low mass transfer rates.}}, \href{https://doi.org/10.48550/arXiv.astro-ph/9509004}{\emph{\aap} {\bfseries 302} (1995) L29} [\href{https://arxiv.org/abs/astro-ph/9509004}{{\ttfamily astro-ph/9509004}}].

\bibitem{mey98wzsge}
E.~{Meyer-Hofmeister}, F.~{Meyer} and B.F.~{Liu}, \emph{{WZ Sagittae - an old dwarf nova}}, \href{https://doi.org/10.48550/arXiv.astro-ph/9809286}{\emph{\aap} {\bfseries 339} (1998) 507} [\href{https://arxiv.org/abs/astro-ph/9809286}{{\ttfamily astro-ph/9809286}}].

\bibitem{mat07wzsgepropeller}
O.M.~{Matthews}, R.~{Speith}, G.A.~{Wynn} and R.G.~{West}, \emph{Magnetically moderated outbursts of {WZ Sagittae}}, {\emph{MNRAS} {\bfseries 375} (2007) 105}.

\bibitem{pat80wzsge}
J.~Patterson, \emph{Rapid oscillations in cataclysmic variables. {IV} -- {WZ Sagittae}}, {\emph{ApJ} {\bfseries 241} (1980) 235}.

\bibitem{wou12ccscl}
P.A.~{Woudt}, B.~{Warner}, A.~{Gulbis}, R.~{Coppejans}, F.-J.~{Hambsch}, A.P.~{Beardmore} et~al., \emph{{CC Sculptoris: a superhumping intermediate polar}}, \href{https://doi.org/10.1111/j.1365-2966.2012.22010.x}{\emph{\mnras} {\bfseries 427} (2012) 1004} [\href{https://arxiv.org/abs/1208.5936}{{\ttfamily 1208.5936}}].

\bibitem{szk17ccsclrzleo}
P.~{Szkody}, A.S.~{Mukadam}, O.~{Toloza}, B.T.~{G{\"a}nsicke}, Z.~{Dai}, A.F.~{Pala} et~al., \emph{{Hubble Space Telescope Ultraviolet Light Curves Reveal Interesting Properties of CC Sculptoris and RZ Leonis}}, \href{https://doi.org/10.3847/1538-3881/aa5c88}{\emph{\aj} {\bfseries 153} (2017) 123} [\href{https://arxiv.org/abs/1702.04076}{{\ttfamily 1702.04076}}].

\bibitem{pav19a19fk}
E.~{Pavlenko}, K.~{Niijima}, P.~{Mason}, N.~{Wells}, A.~{Sosnovskij}, K.~{Antonyuk} et~al., \emph{{ASASSN-18fk: A new WZ Sge-type dwarf nova with multiple rebrightenings and a new candidate for a superhumping intermediate polar}}, {\emph{Contributions of the Astronomical Observatory Skalnate Pleso} {\bfseries 49} (2019) 204} [\href{https://arxiv.org/abs/1907.00623}{{\ttfamily 1907.00623}}].

\bibitem{lop20j2056}
R.~{Lopes de Oliveira}, A.~{Bruch}, C.V.~{Rodrigues}, A.S.~{Oliveira} and K.~{Mukai}, \emph{{CTCV J2056-3014: An X-Ray-faint Intermediate Polar Harboring an Extremely Fast-spinning White Dwarf}}, \href{https://doi.org/10.3847/2041-8213/aba618}{\emph{\apjl} {\bfseries 898} (2020) L40} [\href{https://arxiv.org/abs/2007.13932}{{\ttfamily 2007.13932}}].

\bibitem{gre23v844her}
A.~{Greiveldinger}, P.~{Garnavich}, C.~{Littlefield}, M.R.~{Kennedy}, J.P.~{Halpern}, J.R.~{Thorstensen} et~al., \emph{{A Surprising Periodicity Detected during a Super-outburst of V844 Herculis by TESS}}, \href{https://doi.org/10.3847/1538-4357/acf21b}{\emph{\apj} {\bfseries 955} (2023) 150} [\href{https://arxiv.org/abs/2308.10344}{{\ttfamily 2308.10344}}].

\bibitem{kol25gaia19cwm}
A.I.~{Kolbin}, T.A.~{Fatkhullin}, E.P.~{Pavlenko}, M.V.~{Suslikov}, V.Y.~{Kochkina}, N.V.~{Borisov} et~al., \emph{{Gaia 19cwm{\textemdash}An Eclipsing Dwarf Nova of WZ Sge Type with a Magnetic White Dwarf}}, \href{https://doi.org/10.1134/S1063773725700057}{\emph{Astronomy Letters} {\bfseries 50} (2024) 687} [\href{https://arxiv.org/abs/2502.07447}{{\ttfamily 2502.07447}}].

\bibitem{cas25goto0650}
N.~{Castro Segura}, Z.A.~{Irving}, F.M.~{Vincentelli}, D.~{Altamirano}, Y.~{Tampo}, C.~{Knigge} et~al., \emph{{Bridging the gap: OPTICAM reveals the hidden spin of the WZ Sge star GOTO 065054.49+593624.51}}, \href{https://doi.org/10.1093/mnrasl/slaf038}{\emph{\mnras} {\bfseries 541} (2025) L28} [\href{https://arxiv.org/abs/2501.11669}{{\ttfamily 2501.11669}}].

\bibitem{ver24awdqpo}
M.~{Veresvarska}, S.~{Scaringi}, C.~{Knigge}, J.~{Paice}, D.A.H.~{Buckley}, N.C.~{Segura} et~al., \emph{{Discovery of persistent quasi-periodic oscillations in accreting white dwarfs: a new link to X-ray binaries}}, \href{https://doi.org/10.1093/mnras/stae2279}{\emph{\mnras} {\bfseries 534} (2024) 3087} [\href{https://arxiv.org/abs/2410.01896}{{\ttfamily 2410.01896}}].

\bibitem{tam20j2104}
Y.~{Tampo}, K.~{Naoto}, K.~{Isogai}, T.~{Kato}, M.~{Kimura}, Y.~{Wakamatsu} et~al., \emph{{First detection of two superoutbursts during the rebrightening phase of a WZ Sge-type dwarf nova: TCP J21040470+4631129}}, \href{https://doi.org/10.1093/pasj/psaa043}{\emph{\pasj} {\bfseries 72} (2020) 49} [\href{https://arxiv.org/abs/2004.10508}{{\ttfamily 2004.10508}}].

\bibitem{tam25wzsgetess}
Y.~{Tampo}, N.~{Kojiguchi}, K.~{Isogai}, D.~{Nogami}, H.~{Itoh}, F.-J.~{Hambsch} et~al., \emph{{TESS and ground-based observations of WZ Sge-type dwarf novae in outburst}}, \href{https://doi.org/10.1093/mnras/staf1964}{\emph{\mnras} (2025) } [\href{https://arxiv.org/abs/2511.04175}{{\ttfamily 2511.04175}}].

\bibitem{tov22v455andspec}
G.~{Tovmassian}, B.T.~{G{\"a}nsicke}, J.~{Echevarria}, S.~{Zharikov} and A.~{Ramirez}, \emph{{The Evolution of the Optical Spectrum of V455 Andromedae throughout the 2007 Superoutburst}}, \href{https://doi.org/10.3847/1538-4357/ac930a}{\emph{\apj} {\bfseries 939} (2022) 14}.

\bibitem{rid19j165350}
R.~{Ridden-Harper}, B.E.~{Tucker}, P.~{Garnavich}, A.~{Rest}, S.~{Margheim}, E.J.~{Shaya} et~al., \emph{{Discovery of a new WZ Sagittae-type cataclysmic variable in the Kepler/K2 data}}, \href{https://doi.org/10.1093/mnras/stz2923}{\emph{\mnras} {\bfseries 490} (2019) 5551} [\href{https://arxiv.org/abs/2008.04314}{{\ttfamily 2008.04314}}].

\bibitem{jor24ttisimulation}
L.M.~{Jordan}, D.~{Wehner} and R.~{Kuiper}, \emph{{Two-dimensional simulations of disks in close binaries: Simulating outburst cycles in cataclysmic variables}}, \href{https://doi.org/10.1051/0004-6361/202348726}{\emph{\aap} {\bfseries 689} (2024) A354} [\href{https://arxiv.org/abs/2407.16610}{{\ttfamily 2407.16610}}].

\bibitem{wak17asassn16eg}
Y.~{Wakamatsu}, K.~{Isogai}, M.~{Kimura}, T.~{Kato}, T.~{Vanmunster}, G.~{Stone} et~al., \emph{{ASASSN-16eg: New candidate for a long-period WZ Sge-type dwarf nova}}, \href{https://doi.org/10.1093/pasj/psx094}{\emph{\pasj} {\bfseries 69} (2017) 89} [\href{https://arxiv.org/abs/1708.09206}{{\ttfamily 1708.09206}}].

\bibitem{kim16alcom}
M.~{Kimura}, T.~{Kato}, A.~{Imada}, K.~{Ikuta}, K.~{Isogai}, P.A.~{Dubovsky} et~al., \emph{{Unexpected superoutburst and rebrightening of AL Comae Berenices in 2015}}, \href{https://doi.org/10.1093/pasj/psv121}{\emph{\pasj} {\bfseries 68} (2016) L2} [\href{https://arxiv.org/abs/1511.06596}{{\ttfamily 1511.06596}}].

\bibitem{tam23v627peg}
Y.~{Tampo}, T.~{Kato}, N.~{Kojiguchi}, S.Y.~{Shugarov}, H.~{Itoh}, K.~{Matsumoto} et~al., \emph{{2021 superoutburst of the WZ Sge-type dwarf nova V627 Pegasi lacks an early superhump phase}}, \href{https://doi.org/10.1093/pasj/psad023}{\emph{\pasj} {\bfseries 75} (2023) 619} [\href{https://arxiv.org/abs/2303.17960}{{\ttfamily 2303.17960}}].

\bibitem{kim23j0302}
M.~{Kimura}, K.~{Kashiyama}, T.~{Shigeyama}, Y.~{Tampo}, S.~{Yamada} and T.~{Enoto}, \emph{{MASTER OT J030227.28+191754.5: A Dwarf Nova at a Massive Oxygen-Neon White Dwarf System?}}, \href{https://doi.org/10.3847/1538-4357/acd933}{\emph{\apj} {\bfseries 951} (2023) 124} [\href{https://arxiv.org/abs/2305.15994}{{\ttfamily 2305.15994}}].

\bibitem{tam24j0302}
Y.~{Tampo}, T.~{Kato}, K.~{Isogai}, M.~{Kimura}, N.~{Kojiguchi}, D.~{Nogami} et~al., \emph{{MASTER OT J030227.28+191754.5: An unprecedentedly energetic dwarf nova outburst}}, \href{https://doi.org/10.1093/pasj/psae082}{\emph{\pasj} {\bfseries 76} (2024) 1228} [\href{https://arxiv.org/abs/2408.13783}{{\ttfamily 2408.13783}}].

\bibitem{can88angularmomentumloss}
J.K.~{Cannizzo} and R.E.~{Pudritz}, \emph{{A New Angular Momentum Loss Mechanism for Cataclysmic Variables}}, \href{https://doi.org/10.1086/166241}{\emph{\apj} {\bfseries 327} (1988) 840}.

\bibitem{tet18winddim}
B.E.~{Tetarenko}, J.-P.~{Lasota}, C.O.~{Heinke}, G.~{Dubus} and G.R.~{Sivakoff}, \emph{{Strong disk winds traced throughout outbursts in black-hole X-ray binaries}}, \href{https://doi.org/10.1038/nature25159}{\emph{\nat} {\bfseries 554} (2018) 69} [\href{https://arxiv.org/abs/1801.07203}{{\ttfamily 1801.07203}}].

\bibitem{lon02python}
K.S.~{Long} and C.~{Knigge}, \emph{{Modeling the Spectral Signatures of Accretion Disk Winds: A New Monte Carlo Approach}}, \href{https://doi.org/10.1086/342879}{\emph{\apj} {\bfseries 579} (2002) 725} [\href{https://arxiv.org/abs/astro-ph/0208011}{{\ttfamily astro-ph/0208011}}].

\bibitem{tam21seimeiCVspec}
Y.~{Tampo}, K.~{Isogai}, N.~{Kojiguchi}, H.~{Maehara}, K.~{Taguchi}, T.~{Kato} et~al., \emph{{Spectroscopic and photometric observations of dwarf nova superoutbursts by the 3.8 m telescope Seimei and the Variable Star Network}}, \href{https://doi.org/10.1093/pasj/psab036}{\emph{\pasj} {\bfseries 73} (2021) 753} [\href{https://arxiv.org/abs/2104.04948}{{\ttfamily 2104.04948}}].

\bibitem{kur20seimei}
M.~{Kurita}, M.~{Kino}, F.~{Iwamuro}, K.~{Ohta}, D.~{Nogami}, H.~{Izumiura} et~al., \emph{{The Seimei telescope project and technical developments}}, \href{https://doi.org/10.1093/pasj/psaa036}{\emph{\pasj} {\bfseries 72} (2020) 48}.

\bibitem{tam22v455andspec}
Y.~{Tampo}, D.~{Nogami}, T.~{Kato}, K.~{Ayani}, H.~{Naito}, N.~{Narita} et~al., \emph{{Spectroscopic observations of V455 Andromedae superoutburst in 2007: The most exotic spectral features in dwarf nova outbursts}}, \href{https://doi.org/10.1093/pasj/psac007}{\emph{\pasj} {\bfseries 74} (2022) 460} [\href{https://arxiv.org/abs/2201.09094}{{\ttfamily 2201.09094}}].

\bibitem{tam24dnwind}
Y.~{Tampo}, C.~{Knigge}, K.S.~{Long}, J.H.~{Matthews} and N.C.~{Segura}, \emph{{A disc wind origin for the optical spectra of dwarf novae in outburst}}, \href{https://doi.org/10.1093/mnras/stae1557}{\emph{\mnras} {\bfseries 532} (2024) 1199} [\href{https://arxiv.org/abs/2406.14396}{{\ttfamily 2406.14396}}].

\bibitem{mat25sirocco}
J.H.~{Matthews}, K.S.~{Long}, C.~{Knigge}, S.A.~{Sim}, E.J.~{Parkinson}, N.~{Higginbottom} et~al., \emph{{SIROCCO: a publicly available Monte Carlo ionization and radiative transfer code for astrophysical outflows}}, \href{https://doi.org/10.1093/mnras/stae2677}{\emph{\mnras} {\bfseries 536} (2025) 879} [\href{https://arxiv.org/abs/2410.19908}{{\ttfamily 2410.19908}}].

\bibitem{kat13j1222}
T.~{Kato}, B.~{Monard}, F.-J.~{Hambsch}, S.~{Kiyota} and H.~{Maehara}, \emph{{SSS J122221.7-311523: Double Superoutburst in the Best Candidate for a Period Bouncer}}, \href{https://doi.org/10.1093/pasj/65.5.L11}{\emph{\pasj} {\bfseries 65} (2013) L11} [\href{https://arxiv.org/abs/1307.5936}{{\ttfamily 1307.5936}}].

\bibitem{kim18asassn16dt16hg}
M.~{Kimura}, K.~{Isogai}, T.~{Kato}, K.~{Taguchi}, Y.~{Wakamatsu}, F.-J.~{Hambsch} et~al., \emph{{ASASSN-16dt and ASASSN-16hg: Promising candidate period bouncers}}, \href{https://doi.org/10.1093/pasj/psy037}{\emph{\pasj} {\bfseries 70} (2018) 47} [\href{https://arxiv.org/abs/1803.03179}{{\ttfamily 1803.03179}}].

\bibitem{gre25amcvns}
M.J.~{Green}, J.~{van Roestel} and T.L.S.~{Wong}, \emph{{A catalogue of ultracompact mass-transferring white dwarf binaries}}, \href{https://doi.org/10.1051/0004-6361/202554925}{\emph{\aap} {\bfseries 700} (2025) A107} [\href{https://arxiv.org/abs/2505.10535}{{\ttfamily 2505.10535}}].

\bibitem{iso19nsv1440}
K.~{Isogai}, T.~{Kato}, B.~{Monard}, F.-J.~{Hambsch}, G.~{Myers}, P.~{Starr} et~al., \emph{{NSV 1440: first WZ Sge-type object in AM CVn stars and candidates}}, \href{https://doi.org/10.1093/pasj/psz018}{\emph{\pasj} {\bfseries 71} (2019) 48} [\href{https://arxiv.org/abs/1901.11425}{{\ttfamily 1901.11425}}].

\bibitem{riv22a21au}
L.E.~{Rivera Sandoval}, C.O.~{Heinke}, J.M.~{Hameury}, Y.~{Cavecchi}, T.~{Vanmunster}, T.~{Tordai} et~al., \emph{{The Fast Evolving, Tremendous Blue Superoutburst in ASASSN-21au Reveals a Dichotomy in the Outbursts of Long-period AM CVns}}, \href{https://doi.org/10.3847/1538-4357/ac3fb7}{\emph{\apj} {\bfseries 926} (2022) 10} [\href{https://arxiv.org/abs/2107.11006}{{\ttfamily 2107.11006}}].

\bibitem{tar25j0722}
A.~{Tarasenkov}, K.~{Sokolovsky}, A.~{Dodin}, O.~{Chernyshenko}, S.~{Korotkiy}, I.~{Strakhov} et~al., \emph{{TCP J07222683+6220548: A New AM CVn Type System with Infrequent Outbursts}}, \href{https://doi.org/10.1088/1674-4527/adddec}{\emph{Research in Astronomy and Astrophysics} {\bfseries 25} (2025) 075017} [\href{https://arxiv.org/abs/2505.20842}{{\ttfamily 2505.20842}}].

\bibitem{neu18j1222gwlib}
V.V.~{Neustroev}, K.L.~{Page}, E.~{Kuulkers}, J.P.~{Osborne}, A.P.~{Beardmore}, C.~{Knigge} et~al., \emph{{Superhumps linked to X-ray emission. The superoutbursts of SSS J122221.7-311525 and GW Lib}}, \href{https://doi.org/10.1051/0004-6361/201731719}{\emph{\aap} {\bfseries 611} (2018) A13} [\href{https://arxiv.org/abs/1712.03515}{{\ttfamily 1712.03515}}].

\bibitem{nii21a18ey}
K.~{Niijima}, M.~{Kimura}, Y.~{Wakamatsu}, T.~{Kato}, D.~{Nogami}, K.~{Isogai} et~al., \emph{{Optical Variability Correlated with X-ray Spectral Transition in the Black-Hole Transient ASASSN-18ey = MAXI J1820+070}}, {\emph{VSOLJ\ Variable\ Star\ Bull.} {\bfseries 74} (2021) arXiv:2107.03681} [\href{https://arxiv.org/abs/2107.03681}{{\ttfamily 2107.03681}}].

\bibitem{krz64wzsge}
W.~{Krzeminski} and R.P.~{Kraft}, \emph{{Binary Stars among Cataclysmic Variables. V. Photoelectric and Spectroscopic Observations of the Ultra-Short Binary Nova WZ Sagittae.}}, \href{https://doi.org/10.1086/147995}{\emph{\apj} {\bfseries 140} (1964) 921}.

\bibitem{mun24erositacv}
D.~{Mu{\~n}oz-Giraldo}, B.~{Stelzer} and A.~{Schwope}, \emph{{Cataclysmic variables around the period-bounce: An eROSITA-enhanced multiwavelength catalog}}, \href{https://doi.org/10.1051/0004-6361/202449358}{\emph{\aap} {\bfseries 687} (2024) A305} [\href{https://arxiv.org/abs/2401.17298}{{\ttfamily 2401.17298}}].

\bibitem{ini25sdsscvs}
K.~{Inight}, B.T.~{G{\"a}nsicke}, A.~{Schwope}, S.F.~{Anderson}, E.~{Breedt}, J.R.~{Brownstein} et~al., \emph{{Cataclysmic variables from Sloan Digital Sky Survey - V (2020-2023) identified using machine learning}}, \href{https://doi.org/10.1093/mnras/stae2524}{\emph{\mnras} {\bfseries 536} (2025) 1057} [\href{https://arxiv.org/abs/2406.19459}{{\ttfamily 2406.19459}}].

\bibitem{buc25cvinlsst}
D.A.H.~{Buckley}, Y.~{Tampo}, P.~{Szkody}, M.~{Motsoaledi}, S.~{Scaringi}, M.~{Lochner} et~al., \emph{{Discovering Cataclysmic Variables from the Rubin Observatory LSST}}, \href{https://doi.org/10.3847/1538-4365/ae061b}{\emph{\apjs} {\bfseries 281} (2025) 6} [\href{https://arxiv.org/abs/2509.07298}{{\ttfamily 2509.07298}}].

\bibitem{sca23cvsinlisa}
S.~{Scaringi}, K.~{Breivik}, T.B.~{Littenberg}, C.~{Knigge}, P.J.~{Groot} and M.~{Veresvarska}, \emph{{Cataclysmic variables are a key population of gravitational wave sources for LISA}}, \href{https://doi.org/10.1093/mnrasl/slad093}{\emph{\mnras} {\bfseries 525} (2023) L50} [\href{https://arxiv.org/abs/2307.02553}{{\ttfamily 2307.02553}}].

\bibitem{pog24cvgw}
R.~{Poggiani}, \emph{{Could cataclysmic variables be sources of gravitational waves?}},  in \emph{The Golden Age of Cataclysmic Variables and Related Objects - VI,}, p.~43, Sept., 2024.

\end{thebibliography}\endgroup

\end{document}